# Chiral electric field in relativistic heavy-ion collisions at energies available at the BNL Relativistic Heavy Ion Collider and at the CERN Large Hadron Collider

Yang Zhong (钟洋) [1,2 ;)1]   Chun-Bin Yang (杨纯斌)[1,3]，Xu Cai(蔡勖)[1,3]

Sheng-Qin Feng (冯笙琴)[2,3]

[1]Institute of Particle Physics, Central China Normal University, Wuhan 430079, China

[2]Department of Physics, College of Science, China Three Gorges University，

Yichang 443002, China

[3] Key Laboratory of Quark and Lepton Physics (Huazhong Normal University),

Ministry of Education，Wuhan 430079，China

**Abstract:**   It has been proposed that electric fields may lead to chiral separation in quark-gluon plasma (QGP). This is called the chiral electric separation effect. The strong electromagnetic field and the QCD vacuum can both be completely produced in off-central nuclear-nuclear collision. We use the Woods-Saxon nucleon distribution to calculate the electric field distributions of off-central collisions. The chiral electric field spatial distribution at Relativistic Heavy-Ion Collider (RHIC) and Large Hadron Collider (LHC) energy regions are systematically studied in this paper. The dependence of the electric field produced by the thermal quark in the central position with different impact parameters on the proper time with different collision energies in the RHIC and LHC energy regions are studied in this paper.

## 1   Introduction

Relativistic heavy-ion collisions generate not only hot quark-gluon plasma (QGP) but also large magnetic fields due to the fast motion of the colliding ions [1-6]. Relativistic heavy-ion collisions can also generate strong electric fields due to the event-by-event fluctuation of the positions of the proton in the ions [3,5]. It has been proposed that the electric fields may also lead to chiral separation in the QGP; this is called the chiral electric separation effect (CESE) [7–11].

The origin of electromagnetic field in relativistic heavy-ion collisions comes from collisions

* Supported by National Natural Science Foundation of China (Grants Nos. 11375069, 11435054, 11075061, and 11221504), Excellent Youth Foundation of Hubei Scientific Committee (2006ABB036) and Key Laboratory foundation of Quark and Lepton Physics (Hua-Zhong Normal University)( QLPL2014P01)

 1) Corresponding author: yzhong913@163.com

of two ions of radius *R* with electric charge $Ze$ and extremely high velocity at impact parameter *b*. Lots of analytical and numerical calculations indicate the existence of extremely powerful electromagnetic fields in relativistic heavy-ion collisions [1-5]. It is generally believed that this is the maximum electromagnetic field that nature can produce.

Kharzeev, McLerran and Warringa (KNW) [10] presented a novel mechanism for the study of charge separation of the chiral magnetic effect. The topological charge changing transitions provide the parity (P) and CP violations necessary for charge separation. The variance of the net topological charge change is proportional to the total number of topological charge changing transitions. Therefore if sufficiently hot matter is produced in relativistic heavy-ion collisions, topological charge transitions can take place.

In Refs [12, 13], we used the Woods-Saxon nucleon distribution to replace that of the uniform distribution to improve the magnetic field calculation of the off-central collision based on the KMW theory [10]. The study of the chiral magnetic field distribution at Relativistic Heavy-Ion Collider (RHIC) and Large Hadron Collider (LHC) energy regions has been carried out in Refs. [10, 12, 13]. In this paper, we will calculate the spatial distribution feature of the electric field in the RHIC and LHC energy regions. The electric field distributions of total charges by the thermal quarks produced are also studied in the paper.

The paper is organized as follows. The study of background electric field in relativistic heavy-ion collisions is presented in Section 2. The electric field distributions of total charges by the produced thermal quarks are presented in Section 3. A summary is provided in Section 4.

## 2 Background electric field in relativistic heavy-ion collisions

As the nuclei travel with the speed of light in ultra-relativistic heavy-ion collision experiments, the Lorentz contraction factor $\gamma$ is so large that the two colliding nuclei can be taken as a pancake shape (in the $z = 0$ plane). The Woods-Saxon nuclear distribution is used in Ref. [12, 13] to calculate the magnetic field. The Woods-Saxon nuclear distribution form is as follows:

$$n_A(r) = \frac{n_0}{1 + \exp(\frac{r - R}{d})} , \qquad (1)$$

where $n_0 = 0.17 \, fm^{-3}$, $d = 0.54 \, fm$ and $R = 1.12 A^{1/3} \, fm$. The nuclear charge density of the two-dimensional plane can be given as:

$$\rho_{\pm}(\vec{x}'_{\perp}) = N \cdot \int_{-\infty}^{\infty} dz' \frac{n_0}{1+\exp(\frac{\sqrt{(x' \mp b/2)^2 + y'^2 + z'^2} - R}{d})}, \quad (2)$$

where $N$ is the normalization constant. The number density function $\rho_{\pm}(\vec{x}'_{\perp})$ should be normalized as

$$\int d\vec{x}'_{\perp} \rho_{\pm}(\vec{x}'_{\perp}) = 1. \quad (3)$$

The electric field can be specified in the following way

$$\vec{E} = \vec{E}_p^+ + \vec{E}_p^- + \vec{E}_s^+ + \vec{E}_s^-, \quad (4)$$

where $\vec{E}_p^{\pm}$ and $\vec{E}_s^{\pm}$ are the contributions of the participants and spectators moving in the positive and negative directions, respectively. The contribution of the participants to the electric field is given by

$$\vec{E}_p^{\pm} = \pm Z \alpha_{EM} \int d^2 \vec{x}'_{\perp} \int dY f(Y) \cosh(Y \mp \eta) \rho_{\pm}(\vec{x}'_{\perp}) \theta_{\mp}(\vec{x}'_{\perp}) \frac{\vec{x}'_{\perp} - \vec{x}_{\perp}}{\left[(\vec{x}'_{\perp} - \vec{x}_{\perp})^2 + \tau^2 \sinh^2(Y+\eta)\right]^{3/2}}, \quad (5)$$

where $\theta_{\mp}(\vec{x}'_{\perp}) = \theta\left[R^2 - (\vec{x}'_{\perp} \pm \vec{b}/2)^2\right]$ is a step function, $\eta = \frac{1}{2}\ln[(t+z)/(t-z)]$ is the space-time rapidity and $\tau = (t^2 - z^2)^{1/2}$ is the proper time. The distribution of participants that remain traveling along the beam axis is given by

$$f(Y) = \frac{a}{2\sinh(aY_0)} e^{aY}, \quad -Y_0 \leq Y \leq Y_0 \quad (6)$$

where $a \approx 1/2$ is given by experimental data. The $x$ and $y$ components of $\vec{E}_p^{\pm}$ are given by:

$$\vec{E}_{px}^{\pm} = \pm Z \alpha_{EM} \int d^2 \vec{x}'_{\perp} \int dY f(Y) \cosh(Y \mp \eta) \rho_{\pm}(\vec{x}'_{\perp}) \theta_{\mp}(\vec{x}'_{\perp}) \frac{x' - x}{\left[(\vec{x}' - \vec{x})^2 + \tau^2 \sinh^2(Y+\eta)\right]^{3/2}} \quad (7)$$

$$\vec{E}_{py}^{\pm} = \pm Z \alpha_{EM} \int d^2 \vec{x}'_{\perp} \int dY f(Y) \cosh(Y \mp \eta) \rho_{\pm}(\vec{x}'_{\perp}) \theta_{\mp}(\vec{x}'_{\perp}) \frac{y' - y}{\left[(\vec{x}' - \vec{x})^2 + \tau^2 \sinh^2(Y+\eta)\right]^{3/2}} \quad (8)$$

The contribution of the spectators to the electric field is given by

$$\vec{E}_s^{\pm} = \pm Z \alpha_{EM} \cosh(Y_0 \mp \eta) \int d^2 \vec{x}'_{\perp} \rho_{\pm}(\vec{x}'_{\perp})[1 - \theta_{\mp}(\vec{x}'_{\perp})] \frac{\vec{x}'_{\perp} - \vec{x}_{\perp}}{\left[(\vec{x}'_{\perp} - \vec{x}_{\perp})^2 + \tau^2 \sinh^2(Y+\eta)\right]^{3/2}} \quad (9)$$

The $x$ and $y$ components of $\vec{E}_s^\pm$ are given by:

$$\vec{E}_{sx}^\pm = \pm Z\alpha_{EM} \cosh(Y \mp \eta) \int d^2\vec{x}'_\perp \rho_\pm(\vec{x}'_\perp)[1-\theta_\mp(\vec{x}'_\perp)] \frac{x'-x}{\left[(\vec{x}'-\vec{x})^2 + \tau^2 \sinh^2(Y+\eta)\right]^{3/2}} \quad (10)$$

$$\vec{E}_{sy}^\pm = \pm Z\alpha_{EM} \cosh(Y \mp \eta) \int d^2\vec{x}'_\perp \rho_\pm(\vec{x}'_\perp)[1-\theta_\mp(\vec{x}'_\perp)] \frac{y'-y}{\left[(\vec{x}'-\vec{x})^2 + \tau^2 \sinh^2(Y+\eta)\right]^{3/2}} \quad (11)$$

Here we should explain that the nuclear charge density $\rho$ shown in our paper provides actually only event-averaged distribution of charge nucleons. The actual distribution in a given event is a different form $\rho$.

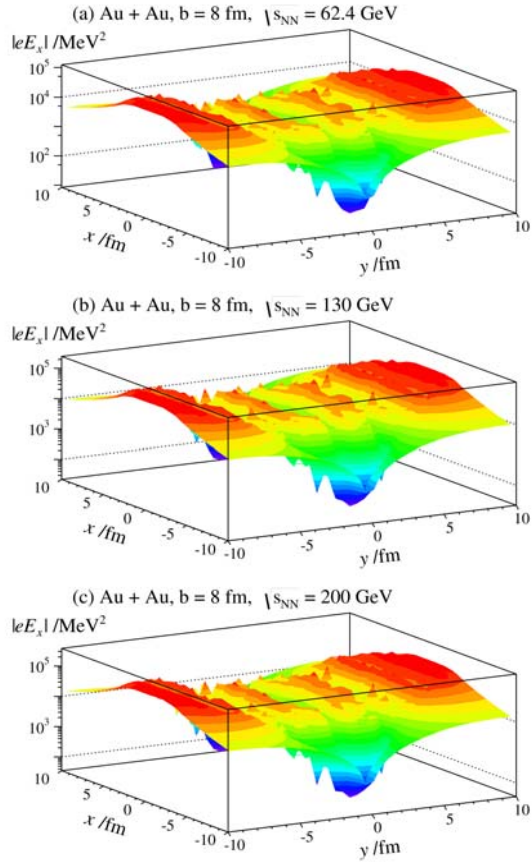

Fig. 1. The dependencies of absolute value of electric field spatial distributions of $|eE_x|$ on different collision energies $\sqrt{s_{NN}}$ = 62.4 GeV(a), 130 GeV(b) and 200 GeV(c), respectively. The impact parameter b = 8 fm and proper time $\tau$ = 0.001 fm.

Figure 1 shows the dependence of electric field absolute value spatial distributions of $|eE_x|$ on different collision energies $\sqrt{s_{NN}}$ = 62.4 GeV, 130 GeV and 200 GeV at proper time $\tau$ = 0.001 fm and impact parameter b = 8 fm. The collision energies shown in Figure 1 are in the

RHIC energy region. The $|eE_x|$ spatial distributions show obvious axis symmetry characteristics along the $x=0$ and $y=0$ axes. There is a valley at the central point ($x=0$, $y=0$) and $|eE_x|$ along the $y = 0$ axis take smaller values. It is interesting to find that $|eE_x|$ increases with the increase of the distance from the $y = 0$ axis.

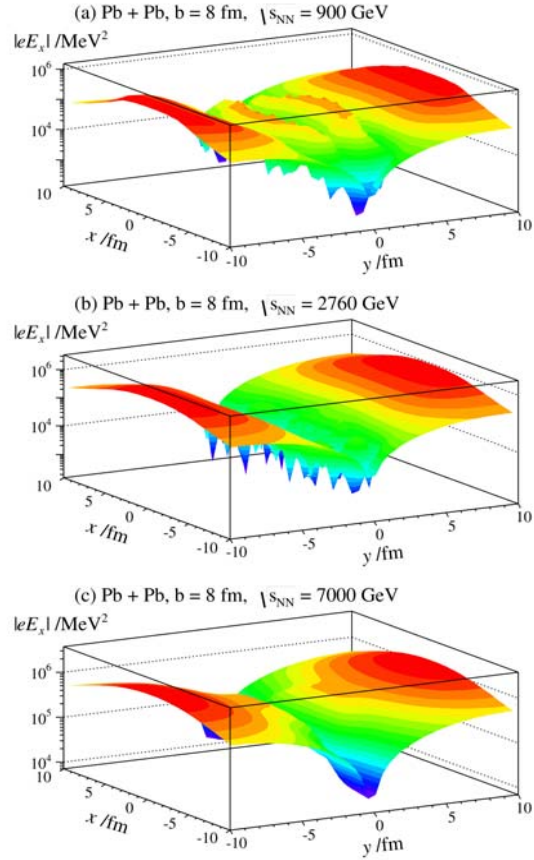

Fig. 2. The dependencies of absolute value of electric field spatial distributions of $|eE_x|$ on different collision energies $\sqrt{s_{NN}}$ = 900 GeV(a), 2760 GeV(b), 7000 GeV(c), respectively. The impact parameters b = 8 fm and proper times $\tau$ = 0.001 fm.

Figure 2 shows the dependence of electric field spatial distributions of $|eE_x|$ on different collision energies $\sqrt{s_{NN}}$ = 900 GeV, 2760 GeV and 7000 GeV at proper time $\tau$ = 0.001 fm and impact parameter b = 8 fm. The collision energies shown in Figure 2 are in the LHC energy region. Comparing with Figure 1, we find that with the increase of collision energy, $|eE_x|$ along the $y = 0$ axis becomes smaller than that in the RHIC energy region. When the CMS energy exceeds 2760 GeV, the original four peaks, which are symmetrically distributed at the two sides of the $y = 0$ axis, are turned into two peaks.

Compared with Fig. 1 and Fig. 2, Fig. 3 shows the dependence of electric field spatial distributions of $|eE_y|$ at different collision energies $\sqrt{s_{NN}}$ = 62.4 GeV, 130 GeV and 200 GeV, at proper time $\tau = 0.001$ fm and impact parameter b = 8 fm. The $|eE_y|$ spatial distributions show obvious axis symmetry characteristics along the $x = 0$ and $y = 0$ axes. There is a deep tunnel along the $x = 0$ axis.

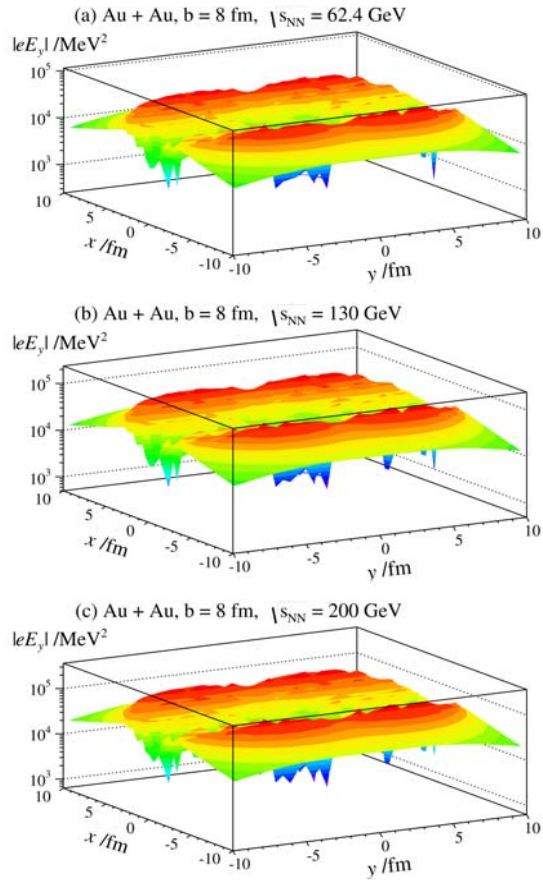

Fig. 3. The dependencies of absolute value of electric field spatial distributions of $|eE_y|$ on different collision energies $\sqrt{s_{NN}}$ = 62.4GeV(a), 130 GeV(b) and 200 GeV(c), respectively. The impact parameter b = 8 fm and proper time $\tau = 0.001$ fm.

Figure 4 shows the dependence of electric field spatial distributions of $|eE_y|$ at different collision energies $\sqrt{s_{NN}}$ = 900 GeV, 2760 GeV and 7000 GeV, at proper time $\tau = 0.001$ fm and impact parameter b = 8 fm. The collision energies shown in Fig. 4 are in the LHC energy region.

Comparing with Fig. 3, we find that with the increasing of collision energy, $|eE_y|$ along the $x = 0$ axis becomes smaller than that in the RHIC energy region. There are two symmetrical platforms on the two sides of the $x = 0$ axis. The $|eE_y|$ spatial distribution is totally different from that of the $|eE_x|$ spatial distributions.

These results for how strong the event-by-event fluctuation of the electric field is are very different from some other publications[3, 4, 5, 7, 9]. These publications [3, 4, 5, 7, 9] used Hijing generator to calculate the event-averaged e<Ex> and e<Ey>. In this paper, we calculate |eEx| and |eEy| by using the methods given by references [10, 11, 12].

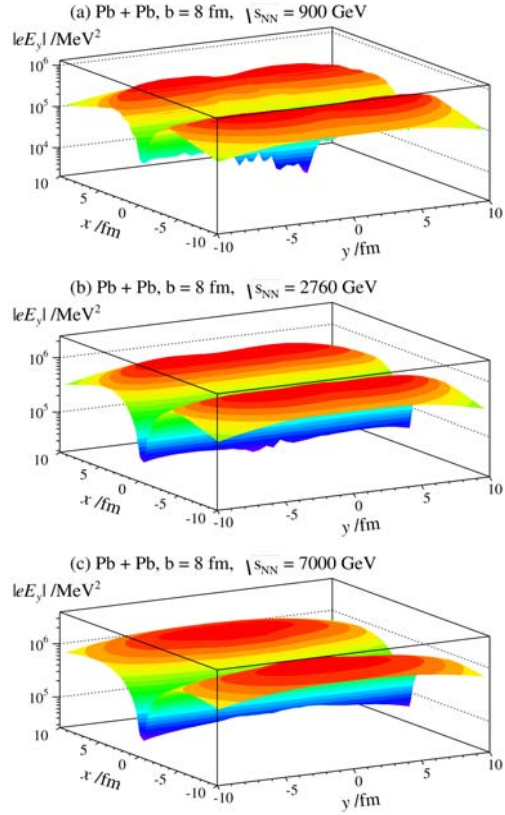

Fig. 4. The dependencies of absolute value of electric field spatial distributions of $|eE_y|$ on different collision energies $\sqrt{s_{NN}}$ = 900GeV(a), 2760 GeV(b) and 7000 GeV(c), respectively. The impact parameter b = 8 fm and proper time $\tau$ = 0.001 fm.

## 3  The electric field of total charges by the produced thermal quarks

First, let us discuss the time and space evolution of the produced thermal quark momentum. According to the theory of the longitudinal collective flow of the relativistic heavy ion collisions, the rapidity distribution of the thermal quarks is [14-21]:

$$f(Y) = K \int_{-Y_{f0}}^{Y_{f0}} (1 + 2\Gamma + 2\Gamma^2) e^{-1/\Gamma} dY_f \tag{12}$$

where K is the normalized constant, and the corresponding function $\Gamma$ is as follows:

$$\Gamma = \frac{T e^{-(\tau-\tau_0)/\tau}}{m \cosh(Y - Y_f)} . \tag{13}$$

In this paper, we used a simple model to study the magnetic field from thermal quarks. We made two assumptions which were used in [12]:

1. We assumed that the contribution to magnetic field from the charge distribution of thermal quarks is approximately proportional to the charge distribution of participant nucleons.

2. We did not distinguish between the charge difference between the u and the d quark, but discussed only one kind of quark with a chemical potential is about 1/3 of the baryon chemical potential.

Based on the Fermi-Dirac statistics [22, 23], the charge distribution function of thermal quarks is given as follows:

$$\rho_{q\pm}(\vec{x}'_\perp) = K_1 \cdot \frac{\rho_\pm(\vec{x}'_\perp)}{1 + \exp\{(\varepsilon - \mu_q)/T\}}, \tag{14}$$

where the nucleon distribution function of the nucleus is:

$$\rho_\pm(\vec{x}'_\perp) = N \cdot \int_{-R}^{R} dz \frac{n_0}{1 + \exp\left(\frac{\sqrt{(\vec{x}' \mp \vec{b}/2)^2 + y^2 + z^2} - R}{d}\right)}, \tag{15}$$

$\varepsilon = \sqrt{m^2 + p_T^2} \cosh Y$ is the thermal quark energy, $K_1$ is the normalized constant, and $\mu_q$ is the quark chemical potential. When $\sqrt{s_{NN}} = 62.4$ GeV, $\mu_q = 1/3 \mu_B = 0.02$ GeV, and when $\sqrt{s_{NN}} = 200$ GeV, $\mu_q = 0.01$ GeV. When $\sqrt{s_{NN}} = 900$, 2760 and 7000 GeV, $\mu_q = 0.005$, 0.002 and 0.001 GeV respectively.

The contribution of the thermal quark to the electric field is:

$$e\bar{E}(\tau,\eta,\vec{x}_\perp) = \pm ZK\alpha_{EM} \int d^2\vec{x}'_\perp \int dY \int_{-Y_{f0}}^{Y_{f0}} dY_f f(Y)(1+2\Gamma+2\Gamma^2)e^{-1/\Gamma}$$
$$\times \cosh(Y \mp \eta)\rho_{q\pm}(\vec{x}'_\perp)\frac{(\vec{x}'_\perp - \vec{x}_\perp)}{\left[(\vec{x}'_\perp - \vec{x}_\perp)^2 + \tau^2 \sinh(Y_0 \mp \eta)^2\right]^{3/2}} \quad (16)$$

where proper time is $\tau = \sqrt{t^2 - z^2}$, and pseudo-rapidity is $\eta = \frac{1}{2}\log[(t+z)/(t-z)]$。

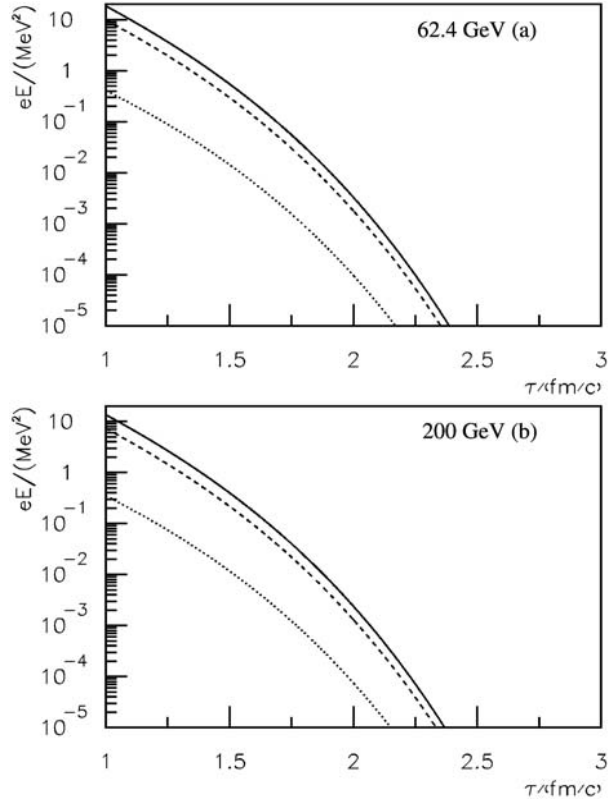

Fig. 5 The dependence of the electric field $eE$ produced by the thermal quark in the central region with different impact parameters on the proper time $\tau$ with $\sqrt{s_{NN}}$ = 62.4 GeV and 200 GeV in the RHIC energy region. The real line is for b = 4 fm, the dashed line is for b = 8 fm and the dotted line is for b =12 fm.

Figure 5 shows the dependence of the electric field $eE$ produced by the thermal quark in the central region with different impact parameters on the proper time $\tau$ with $\sqrt{s_{NN}}$ = 62.4 GeV and 200GeV in the RHIC energy region. It is well known that the formation time of quark gluon plasma (QGP) is $\tau \approx 1$ fm, so the thermal quark evolution time should be larger than 1 fm. From Fig. 5 (a), the maximum of the electric field $eE$ at $\sqrt{s_{NN}}$ = 62.4 GeV is about 20 MeV$^2$, which is much smaller than that of the electric field of nuclear collisions. For the periphery collisions, the contribution of produced thermal quarks to the electric field $eE$ becomes very small. The electric field formed at the beginning of the collision is only 0.4 MeV² in the case $b = 12\,fm$. At the same time, we find that the electric field decays rapidly with the time evolution. Even in the case of $b = 4\,fm$ collision, the electric field is quickly reduced to about $10^{-5}$ MeV². Figure 5 (b) shows the same as Fig. 5(a) but for $\sqrt{s}$ = 200 GeV. There is almost the same variation of the electric field with the time $\tau$ as in Fig. 5 (a).

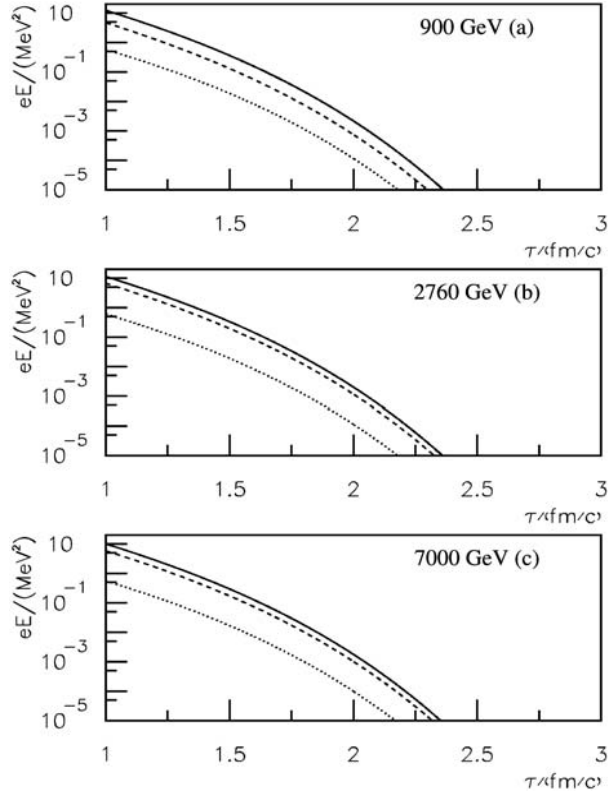

Figure 6 The dependence of the electric field $eE$ produced by the thermal quark in the central region with different impact parameters on the proper time $\tau$ with $\sqrt{s_{NN}}$ = 900 GeV, 2760 GeV and 7000 GeV in the LHC energy region. The real line is for b = 4 fm, the dashed line is for b = 8 fm and the dotted line is for b =12 fm.

Figure 6 shows the dependence of the electric field $eE$ produced by the thermal quarks in the central region with different impact parameters on the proper time $\tau$ with $\sqrt{s_{NN}}$ = 900 GeV, 2760 GeV and 7000 GeV in the LHC energy region. From Figure 6, one can find that the electric field $eE$ increases with the decrease of the impact parameter $b$. The maximum of the electric field $eE$ at $\sqrt{s_{NN}}$ = 900 GeV is about 10 MeV². This value is slightly smaller than that of 20 MeV² at $\sqrt{s_{NN}}$ =62.4 GeV in the RHIC energy region.

It is found that the electric field decays rapidly with the time evolution. Even in the case of $b = 4\,fm$ collisions, the electric field is quickly reduced to about $10^{-3}$ MeV$^2$. Figure 6 (b, c) shows the same as Fig. 6(a) but for $\sqrt{s_{NN}}$ = 2760 GeV and 7000 GeV. It is found that there is almost the same variation of the electric field with time $\tau$ as as Fig. 6(a).

## 4 Summary and conclusions

It is shown that an enormous electric field can indeed be created in off-central heavy-ion collisions. The electric field distributions of $|eE_y|$ and $|eE_x|$ are highly inhomogeneous, like the magnetic field distributions. The enormous electric field is produced just after the collision, and the magnitude of electric field of the LHC energy region is larger than that of the RHIC energy region at small proper time. These highly inhomogeneous distribution features of electric field in RHIC and LHC energy regions will help us to study the experimental results given by RHIC and LHC.

The dependencies of electric field spatial distributions of $|eE_y|$ and $|eE_x|$ on different collision energies $\sqrt{s_{NN}}$ = 900 GeV, 2760 GeV and 7000 GeV at LHC energy region and at $\sqrt{s_{NN}}$ = 62.4 GeV, 130 GeV and 200 GeV at RHIC energy region proper time $\tau$ = 0.001 fm and impact parameter b = 8 fm are studied in this paper. The collision energies mentioned in this paper covered the whole RHIC and LHC energy regions. The feature $|eE_y|$ spatial distribution is totally different from that of $|eE_x|$ spatial distributions. There are two symmetrical platforms as shown for $|eE_y|$ on the two sides of the $x$ = 0 axis. It is found that the $|eE_x|$ takes small values along the $y$ = 0 axis, and increases with the increase of the distance from the $y$ = 0 axis.

We also study the dependence of the electric field $eE$ produced by the thermal quark in the central region with different impact parameters on the proper time $\tau$ in the RHIC and LHC energy region. One can find that the electric field produced by thermal quarks is much smaller than that of nuclear collisions. The maximum of the electric field $eE$ at $\sqrt{s_{NN}}$ = 900 GeV is about $10$ MeV$^2$. This value is slightly smaller than that of 20 MeV$^2$ at $\sqrt{s_{NN}}$ =62.4 GeV in the

RHIC energy region. For the periphery collisions, the contribution of produced particles to the electric field $eE$ becomes very small.

**References**

[1] G. L. Ma, and X. G. Huang. Phys. Rev. C, **91:** 054901 (2015)

[2] J. Bloczynski, X. G. Huang, X. Zhang , and J. Liao, Phys. Lett. B, **718:** 1529 (2013)

[3] V. Skokov, A. Y. Illarionov, and V. Toneev, Int. J. Mod. Phys. A, **24:** 5925 (2009)

[4] A. Bzdak and V. Skokov, Phys. Lett. B, **710:** 171 (2012)

[5] W. T. Deng, and X. G. Huang, Phys. Rev. C, **85:** 044907 (2012)

[6 ] V. Voronyuk, V. D. Toneev, W. Cassing, E. L. Bratkovskaya, V. P. Konchakovski, and S. A. Voloshin, Phys. Rev. C, **83**: 054911 (2011)

[7] X. G. Huang, and J. Liao, Phys. Rev. Lett. **110**: 232302 (2013)

[8] Y. Jiang, X. G. Huang, and J. Liao, Phys. Rev. D, **91**: 045001 (2015)

[9] S. Pu, S. Y. Wu, and D. L. Yang,  Phys. Rev. D, **89**: 085024 (2014)

[10] D. E. Kharzeev, L. D. McLerran, H. J. Warringa, Nucl. Phys. A, **803**:227 (2008)

[11] K .Tuchin, Advances in High Energy Physics, **2013**: 490495 (2013)

[12] Y. J. Mo, S. Q. Feng, and Y. F. Shi,   Phys. Rev. C, **88**: 024901 (2013)

[13] Y. Zhong, C. B. Yang, X. Cai, and S. Q. Feng , Advances in High Energy Physics, **2014**:193039 (2014)

[14] S. Q. Feng, L. Feng, and L. S. Liu, Phys. Rev. C, **63**: 014901（2000）

[15] S. Q. Feng, Y. Zhong, Phys. Rev. C, **83**: 034908 (2011)

[16] S. Q. Feng, W. Xiong, Phys. Rev. C, **77**: 044906 (2008)

[17] E. Schnedermann, J. Sollfrank, U. Heinz, Phys. Rev. C, **48**:2462 (1993)

[18] E. Schnedermann, J. Sollfrank, U. Heinz, Prog. Part. Nucl. Phys, **30**: 401 (1993)

[19] X. Cai, S. Q. Feng, Y. D. Li, C. B. Yang,   and D. C. Zhou, Phys. Rev. C, **51:** 3336 (1995)

[20] P. Braun-Munzinger, J. Stachel, J. P. Wessels, and X. Xu, Phys. Lett. B, **344:** 43 (1995)

[21] P. Braun-Munzinger, I. Heppe, and J. Stachel, Phys. Lett. B, **465**:15 (1999)

[22] F. Becattini, J. Cleymans, A. Keranen, E. Suhonen, and K. Redlich, Phys. Rev. C, **64**: 024901 (2001)

[23] J. Cleymans, K. Redlich, Phys. Rev. Lett. **81:** 5284 (1998)